\documentclass[12pt,a4paper]{article}

\usepackage{amssymb}
\usepackage{amsmath}
\usepackage{amsfonts}
\usepackage{authblk}
\usepackage{setspace}

\numberwithin{equation}{section}

\newcommand{\BbbZ}{\mathbb{Z}}
\newcommand{\BbbC}{\mathbb{C}}

\DeclareMathOperator{\diag}{diag}

\title{Near-Dirichlet quantum dynamics for a\\ 
$p^3$-corrected particle on an interval}

\author{Jorma Louko}
\affil{School of Mathematical Sciences, 
University of Nottingham,\\ 
Nottingham NG7 2RD, 
UK}

\date{{\small April 2014, revised March 2015. 
Published in Gen.\ Relativ.\ Gravit.\ {\bf 47}, 55 (2015).}}

\begin{document}

\maketitle
\begin{abstract}
We study a nonrelativistic quantum mechanical 
particle on an interval of finite length 
with a Hamiltonian that has a $p^3$ correction term, modelling potential 
low energy quantum gravity effects. 
We describe explicitly the $U(3)$ family of the 
self-adjoint extensions of the 
Hamiltonian and discuss several subfamilies of interest. 
As the main result, we find a family of self-adjoint Hamiltonians, 
indexed by four continuous parameters and one binary parameter, 
whose spectrum and eigenfunctions are perturbatively 
close to those of the 
uncorrected particle with Dirichlet boundary conditions, 
even though the Dirichlet condition as 
such is not in the $U(3)$ family. 
Our boundary conditions do not single out distinguished 
discrete values for the 
length of the interval in terms of the underlying quantum gravity scale. 
\end{abstract}


\section{Introduction}

Several theories of quantum spacetime suggest that low energy
corrections due to quantum gravity can be modelled by adding to the
conventional quantum mechanical position or momentum operators terms
that depend on higher powers of the
momentum~\cite{AmelinoCamelia:2008qg}. Such corrections could be
experimentally accessible at low energies through their effects on the
the spectra of quantum mechanical observables, or through their
effects on uncertainty relations. An overview can be found in
\cite{AmelinoCamelia:2008qg}. A~case study with a specific form of
the correction terms is given in~\cite{Kempf:1994su}. 
A~discussion within quantum field theory is given
in~\cite{Husain:2012im}. 

In this paper we consider the quantum mechanics of a nonrelativistic
particle with a $p^3$ correction on an interval. The physical
motivation to work on an interval of finite length, rather than on the
full real line, is to relate the $p^3$ term to ideas about
discreteness of spacetime: might the coefficient of the $p^3$
correction, of quantum gravitational origin, single out some discrete
values of the interval's length as physically
preferred~\cite{Ali:2009zq}?

A technical issue on the interval is that writing down the Hamiltonian
as a differential operator, with or without the $p^3$ term, does not
suffice to define a quantum theory with unitary evolution. What is
required is to specify at the ends of the interval boundary conditions
that define the Hamiltonian as a self-adjoint operator
\cite{reed-simonII,blabk,Bonneau:1999zq}.  Without the $p^3$ term, the
allowed boundary conditions form a $U(2)$ family, which includes as
special cases Dirichlet, Neumann and Robin boundary conditions at each
of the two ends, but also boundary conditions that relate the two
ends, including periodic boundary
conditions~\cite{Bonneau:1999zq,Asorey:2004kk,Ibort:2013ab,Asorey:2013wca,Asorey:2013wvh,Munoz-Castaneda:2014yea}.
With the $p^3$ term, by contrast, the allowed boundary conditions form
a $U(3)$ family~\cite{Balasubramanian:2014pba}, within which the
uncorrected $U(2)$ family is embedded in a rather suble way, as we
shall show.  In particular, the $U(3)$ family does not contain the
Dirichlet conditions of the uncorrected theory.  Yet it is the
Dirichlet conditions that can be regarded as generic in the
uncorrected theory, as they tend to ensue when the finite interval is
built as the limit of a confining potential without
fine-tuning~\cite{walton}.

The following question hence arises. We wish to view the $p^3$ term as
a small correction. Given a choice of boundary conditions within the
uncorrected $U(2)$ family, do there exist boundary conditions in the
corrected $U(3)$ family for which the effects of the $p^3$ term remain
small, in the sense of perturbative expandability~\cite{Simon:1990ic}:
do the corrected eigenenergies and eigenfunctions approach the 
uncorrected ones when the coefficient of the $p^3$ term goes to zero?

The main result of this paper is to show that the answer is
affirmative for the Dirichlet boundary conditions of the uncorrected
theory, and that the subfamily of $U(3)$ for which this happens is,
under certain technical assumptions, indexed by four continuous
parameters and one binary parameter. For this subfamily of~$U(3)$, the
spectrum of the $p^3$-corrected theory does however not appear to have
structure that would single out distinguished discrete values of the
interval's length in terms of the coefficient of the $p^3$ term.

As an intermediate result, we give an explicit 
description of the full $U(3)$ family of
boundary conditions in the $p^3$-corrected theory.  The $U(3)$ family is in
particular seen to contain the $U(1)$ subfamily of periodicity up to a
prescribed phase. Within this $U(1)$ subfamily the $p^3$
corrections are small in the sense of perturbative expandability, 
and while this smallness is not uniform over the full set of eigenvalues, 
we show that the time evolution operator of the corrected theory converges 
to that of the uncorrected theory in the strong operator topology. 

We also show that in the $p^3$-corrected theory, boundary conditions
independent of the second derivative of the wave function form a
$U(1)$ subfamily in which the wave function vanishes at both ends and
its first derivative is periodic up to a prescribed phase. Numerical
evidence suggests that the eigenenergies within this $U(1)$ subfamily
approach the eigenenergies of the uncorrected Dirichlet theory as the
coefficient of the $p^3$ term approaches zero; however, an analytic
argument shows that derivatives of the $p^3$-corrected eigenfunctions
cannot approach those of the uncorrected Dirichlet theory. The
closeness in the eigenenergies with these boundary conditions does
hence not extend to closeness in all quantum mechanical observables,
in particular in observables involving derivatives.

We begin by introducing in Section 
\ref{sec:Hamiltonian-and-extensions}
the Hamiltonian and writing down its $U(3)$ family of self-adjoint extensions, 
deferring technical material to two appendices. 
Relevant facts about the uncorrected Hamiltonian are 
collected in Section~\ref{sec:uncorrected-Hamiltonian}. 
Section \ref{sec:special-Uone-families} discusses two special 
$U(1)$ subfamilies of boundary conditions, 
first periodicity up to a prescribed phase, and then conditions 
independent of the second derivative. 
The main results about perturbatively near-Dirichlet boundary 
conditions are given in Section~\ref{sec:pert-spectrum}. 
Section \ref{sec:conclusions} presents 
a brief summary and concluding remarks. 

We maintain physical units in that the length of the interval has the
physical dimension of length. We however drop an overall
multiplicative constant from the Hamiltonian so that energies
have units of inverse length squared and reduced Planck's constant 
$\hbar$ has units of inverse length. 
The superscript star (${}^*$) denotes complex conjugation. 
The superscript dagger (${}^\dagger$) denotes the  
Hermitian conjugate on matrices and the adjoint on operators.

\section{$p^3$-corrected Hamiltonian and its 
self-adjoint extensions\label{sec:Hamiltonian-and-extensions}}

We work in the Hilbert space ${\mathcal{H}} = L_2([0,L], dx)$, 
where $L$ is a positive constant with the physical dimension of length. 
We consider in ${\mathcal{H}}$ the Hamiltonian operator 
\begin{align}
H = - \partial_x^2 - i q L \partial_x^3 
\ , 
\label{eq:Ham-def}
\end{align}
where $q$ is a dimensionless positive constant. 

For $q=0$, $H$ reduces to the Hamiltonian of a free nonrelativistic
particle. The term that involves $q$ can be thought of as an
effective quantum gravity correction, proportional to
$p^3$~\cite{AmelinoCamelia:2008qg,Ali:2009zq}.  It would be possible
to scale $L$ out of the problem by writing $x = Ly$, where $0\le y \le
1$, and working in the Hilbert space $L_2([0,1], dy)$, 
but we prefer to keep
$L$ in the formulas, in view of potential applications to the
underlying quantum gravity context.  Note that assuming $q>0$ is no
loss of generality since the sign of $q$ can be changed by the
reparametrisation $x \to L-x$.

We take the domain of $H$ to be initially
$\mathcal{C}^{\infty}_{c}(0,L)$.  $H$ is then densely defined and
symmetric.  As $H$ and its adjoint $H^\dagger$ are third-order
differential operators, the solutions to $H^\dagger\psi = \pm i \psi$
are square integrable on $[0,L]$ and form a three-dimensional vector
space for each sign. It follows from von Neumann's theorem that the
self-adjoint extensions of $H$ form a $U(3)$ family
\cite{reed-simonII,blabk,Bonneau:1999zq,Balasubramanian:2014pba}.

To write down the boundary condition that specifies the self-adjoint
extensions of~$H$, we note that if $\psi$ and $\phi$ are smooth
functions on $[0,L]$, the condition $(\psi, H \phi) = (H\psi, \phi)$
can be written as $C(u,v) =0$, where
\begin{align}
C(u,v) := u^\dagger
\begin{pmatrix}
G& 0 \\
0 & -G  
\end{pmatrix} v 
\ , 
\label{eq:Cform-def-maintext}
\end{align}
\begin{align}
&G  := 
\begin{pmatrix}
0 & i & -q \\
-i & q & 0 \\
-q & 0 & 0 \\
\end{pmatrix}
\ , 
\label{eq:Gmatrix-cubic}
\end{align}
and 
\begin{align}
u = 
\begin{pmatrix}
\psi(0)\\
L \psi'(0)\\
L^2 \psi''(0)\\
\psi(L)\\
L \psi'(L)\\
L^2 \psi''(L)\\
\end{pmatrix}
\ , 
\ \ 
v = 
\begin{pmatrix}
\phi(0)\\
L \phi'(0)\\
L^2 \phi''(0)\\
\phi(L)\\
L \phi'(L)\\
L^2 \phi''(L)\\
\end{pmatrix}
\ .  
\label{eq:uvlong-def-maintext}
\end{align} 
In terms of
\eqref{eq:Cform-def-maintext}--\eqref{eq:uvlong-def-maintext}, the
self-adjointness conditions for $H$ are the maximal linear subspaces
of $\BbbC^6$ on which the sesquilinear form
\eqref{eq:Cform-def-maintext} vanishes
\cite{reed-simonII,blabk,Bonneau:1999zq}.  These subspaces are found
in Appendix~\ref{app:subspaces}. We collect here the outcome.

The matrix $G$ \eqref{eq:Gmatrix-cubic} is Hermitian, and its
characteristic polynomial is the cubic 
\begin{align}
P_G(\lambda) = -
\lambda^3+q\lambda^2+(1+ q^2)\lambda-q^3
\ . 
\label{eq:charpol}
\end{align}
$G$~has three distinct
eigenvalues, which we denote in increasing order by $\lambda_-$,
$\lambda_0$ and~$\lambda_+$, and it can be shown that $\lambda_-< 0 <
\lambda_0 < q < \lambda_+$. Let
\begin{align}
\begin{pmatrix}
a_1& a_2 & a_3
\end{pmatrix} 
\ , 
\ \ 
\begin{pmatrix}
b_1& b_2 & b_3
\end{pmatrix} 
\ , 
\ \ 
\begin{pmatrix}
c_1& c_2 & c_3
\end{pmatrix} 
\ , 
\label{eq:eigen-covectors} 
\end{align}
be normalised eigen-covectors for respectively 
$\lambda_-$, $\lambda_+$ and~$\lambda_0$, and let 
\begin{align}
\begin{pmatrix}
\rho_1\\
\rho_2\\
\rho_3\\
\end{pmatrix}
=
\begin{pmatrix}
\phi(0)\\
L \phi'(0)\\
L^2 \phi''(0)\\
\end{pmatrix}
\ , 
\ \ 
\begin{pmatrix}
\sigma_1\\
\sigma_2\\
\sigma_3\\
\end{pmatrix}
=
\begin{pmatrix}
\phi(L)\\
L \phi'(L)\\
L^2 \phi''(L)\\
\end{pmatrix}
\ . 
\label{eq:3by3elements}
\end{align} 
The self-adjointness boundary conditions for $H$ then read 
\begin{align}
U 
\begin{pmatrix}
\sqrt{-\lambda_-} \left( a_1 \sigma_1 + a_2 \sigma_2 + a_3 \sigma_3\right) \\
\sqrt{\lambda_+} \left( b_1 \rho_1 + b_2 \rho_2 + b_3 \rho_3\right) \\
\sqrt{\lambda_0} \left( c_1 \rho_1 + c_2 \rho_2 + c_3 \rho_3\right) \\
\end{pmatrix} 
= 
\begin{pmatrix}
\sqrt{-\lambda_-} \left( a_1 \rho_1 + a_2 \rho_2 + a_3 \rho_3\right) \\
\sqrt{\lambda_+} \left( b_1 \sigma_1 + b_2 \sigma_2 + b_3 \sigma_3\right) \\
\sqrt{\lambda_0} \left( c_1 \sigma_1 + c_2 \sigma_2 + c_3 \sigma_3\right) \\
\end{pmatrix}
\ , 
\label{eq:3by3condition}
\end{align}
where the matrix $U \in U(3)$ specifies the self-adjoint extension. 

Three remarks are in order. First, the notation for the
eigen-covectors \eqref{eq:eigen-covectors} is chosen to avoid
complex conjugates in~\eqref{eq:3by3condition}. 
The corresponding eigenvectors are the
Hermitian conjugates of \eqref{eq:eigen-covectors} but they 
will not be needed. 

Second, the explicit expressions for the eigenvalues from the cubic
solution formula are cumbersome, 
but it can be verified using \eqref{eq:charpol}
that if $\lambda$ is an eigenvalue, the corresponding
eigen-covector is proportional to 
\begin{align}
\begin{pmatrix}
\lambda(\lambda-q) &  i\lambda & q(q-\lambda)
\end{pmatrix} 
\ . 
\label{eq:eigencovector-explicit}
\end{align}
Using \eqref{eq:charpol} and \eqref{eq:eigencovector-explicit} 
allows us to verify by polynomial algebra identities 
that will be needed in
subsection~\ref{subsec:bc-nophipp}, including
\begin{align}
\lambda_- |a_3|^2 + \lambda_+ |b_3|^2 + \lambda_0 |c_3|^2 = 0
\ . 
\label{eq:ev-identity1}
\end{align}

Third, the eigenvalues and the eigen-covectors have small $q$
expansions in non-negative integer powers of~$q$. These
expansions are collected in Appendix~\ref{app:expansion}.

\section{Uncorrected Hamiltonian\label{sec:uncorrected-Hamiltonian}}

In the limit $q\to0$, $H$ \eqref{eq:Ham-def} becomes 
\begin{align}
H^{q=0} = - \partial_x^2 
\ , 
\label{eq:Ham-q=0-def}
\end{align}
and from the small $q$ expansions of the eigenvalues and
eigen-covectors of $G$ given in Appendix \ref{app:expansion} it is
seen that the boundary condition \eqref{eq:3by3condition} reduces to
\begin{align}
U_2
\begin{pmatrix}
\phi(L)  - i L \phi'(L)  \\
\phi(0)  + i L \phi'(0) \\
\end{pmatrix} 
= 
\begin{pmatrix}
\phi(0)  - i L \phi'(0) \\
\phi(L)  + i L \phi'(L) \\
\end{pmatrix}
\ , 
\label{eq:q=0-condition}
\end{align}
where $U_2 \in U(2)$. 
$H^{q=0}$ is the Hamiltonian of the free nonrelativistic particle, 
and \eqref{eq:q=0-condition} is its well-known $U(2)$ family 
of self-adjointness conditions on the 
interval~\cite{Bonneau:1999zq,Asorey:2004kk,Ibort:2013ab,Asorey:2013wca,Asorey:2013wvh,Munoz-Castaneda:2014yea}. 

Our main interest is in the choice 
$U_2 = \bigl(
\begin{smallmatrix}
0 & -1\\
-1 & 0
\end{smallmatrix}
\bigr)$, which 
gives the Dirichlet boundary condition, 
\begin{align}
\phi(0) = \phi(L) = 0 
\ . 
\label{eq:q=0-Dirichlet}
\end{align}
The eigenfunctions are poportional to 
\begin{align}
\sin(m \pi x/L)
\ , 
\ \ 
m = 1, 2, \ldots
\ , 
\label{eq:q=0-Dir-efunctions}
\end{align} 
and the eigenenergies are 
\begin{align}
E_m^{q=0, \text{Dirichlet}} = m^2 \pi^2 L^{-2}
\ , 
\ \ \ 
m = 1, 2, \ldots
\ . 
\label{eq:q=0-Dirichlet-eigen}
\end{align}

We will also be interested in the 
$U(1)$ family of extensions in which $U_2 = \bigl(
\begin{smallmatrix}
e^{-i\beta} & 0\\
0 & e^{i\beta}
\end{smallmatrix}
\bigr)$, 
where $0\le \beta < 2\pi$. The boundary condition is 
\begin{align}
\phi(L) = e^{i\beta} \phi(0)
\ , 
\ \ 
\phi'(L) = e^{i\beta} \phi'(0)
\ , 
\label{eq:periodic-q=0-bcs}
\end{align}
which means that the eigenfunctions are periodic 
up to the prescribed phase~$e^{i\beta}$. 
The eigenfunctions are proportional to 
$\exp\bigl(i(2\pi m +\beta)x/L\bigr)$, 
where $m \in \BbbZ$, and the eigenenergies are 
\begin{align}
E_m^{q=0, \beta} = {(2\pi m +\beta)}^2 L^{-2}
\ , \ \ m \in \BbbZ
\ . 
\label{eq:q=0-periodic-eigen}
\end{align}

We note that while the Dirichlet spectrum
\eqref{eq:q=0-Dirichlet-eigen} is positive definite, and the spectrum
\eqref{eq:q=0-periodic-eigen} is positive definite for $\beta\ne0$ and
positive semidefinite for $\beta=0$, there exist boundary conditions
for which the spectrum is not positive definite, and the ground state
energy can be made arbitrarily negative. As an example, consider the 
$U(1)$ family of extensions in which 
$U_2 = \bigl(
\begin{smallmatrix}
0 & - e^{-i\gamma}\\
- e^{-i\gamma} & 0 
\end{smallmatrix}
\bigr)$, 
where $0\le \gamma < 2\pi$. The boundary condition is 
\begin{align}
\cos(\gamma/2)\phi(0) = - \sin(\gamma/2) L \phi'(0)
\ , 
\ \ 
\cos(\gamma/2)\phi(L) = \sin(\gamma/2) L \phi'(L)
\ , 
\label{eq:neardir-q=0-bcs}
\end{align}
which includes the Dirichlet condition \eqref{eq:q=0-Dirichlet} 
as the special case $\gamma=0$. 
When $\gamma=0$ or $\pi \le \gamma <2\pi$, 
there are no negative eigenenergies. 
However, when $0 < \gamma < \pi$, 
there is a negative energy ground state, and when 
$0 < \gamma < 2\arctan(\tfrac12)$, there is also one 
negative energy excited state: 
the respective eigenenergies $E_0$ and $E_1$ 
are obtained as the unique negative solutions to
\begin{subequations}
\begin{align}
\tan(\gamma/2) \sqrt{- E_0 L^2} 
&= \coth\!\left(\tfrac12 \sqrt{- E_0 L^2} \, \right)
\ , 
\\
\tan(\gamma/2) \sqrt{- E_1 L^2} 
&= \tanh\!\left(\tfrac12 \sqrt{- E_1 L^2} \, \right)
\ . 
\end{align}
\end{subequations}
In the limit $\gamma\to0_+$, 
the two negative eigenenergies disappear by 
descending to negative infinity, 
while the rest of the spectrum approaches 
the Dirichlet spectrum~\eqref{eq:q=0-Dirichlet-eigen}.

\section{Two special $U(1)$ boundary condition 
families\label{sec:special-Uone-families}}

In this section we consider two special $U(1)$ boundary condition families. 
The first family is manifestly not close to 
the Dirichlet condition of the unperturbed theory: 
instead, it extends the periodicity up to a prescribed phase 
\eqref{eq:periodic-q=0-bcs} to the perturbed theory, 
and its purpose is to provide 
an explicitly solvable example in which both the quantitative 
and qualitative features of the $q\to0$ limit can be analysed. 
In particular, both the eigenenergies and the eigenstates 
will be seen to be perturbatively close to those of the unperturbed theory, 
and the uniformness of this closeness can be characterised in terms 
of the topology in which the time evolution operator of the 
perturbed theory converges to that of the unperturbed theory. 
The purpose of the second family is to show that any boundary 
condition in which both the eigenenergies and the eigenstates 
are perturbatively close to those of the unperturbed Dirichlet  theory 
must necessarily involve conditions on the second derivative 
of the wave function.

\subsection{Periodicity up to a prescribed 
phase\label{subsec:q=0:per-upto-phase}}

Consider in \eqref{eq:3by3condition} the choice 
$U = \diag\bigl(e^{-i\beta}, e^{i\beta}, e^{i\beta}\bigr)$, where 
$0\le \beta < 2\pi$. As the eigen-covectors \eqref{eq:eigen-covectors} 
are linearly independent, 
\eqref{eq:3by3condition} is equivalent to 
\begin{align}
\phi(L) = e^{i\beta} \phi(0)
\ , 
\ \ 
\phi'(L) = e^{i\beta} \phi'(0)
\ , 
\ \ 
\phi''(L) = e^{i\beta} \phi''(0)
\ , 
\label{eq:periodic-q-bcs}
\end{align}
which means that the eigenfunctions are periodic 
up to the prescribed phase~$e^{i\beta}$. 
The eigenfunctions are proportional to 
$\exp\bigl(i(2\pi m +\beta)x/L\bigr)$, 
where $m \in \BbbZ$, and the eigenenergies are 
\begin{align}
E_m^{\beta} = {(2\pi m +\beta)}^2 
\bigl(1 - q (2\pi m +\beta) \bigr)
L^{-2}
\ , \ \ m \in \BbbZ
\ . 
\label{eq:periodic-eigen}
\end{align}

From this explicit solution we can make the following three observations. 

First, for given~$m$, the eigenenergy $E_m^{\beta}$ \eqref{eq:periodic-eigen} 
converges to that of the unperturbed theory \eqref{eq:q=0-periodic-eigen}
as $q\to0$. Also, the corresponding eigenfunction and all of 
its derivatives converge to those of the unperturbed theory, 
in (say) the $L_2$ norm. 
The $q>0$ theory
with the boundary condition 
\eqref{eq:periodic-q-bcs} is in this sense perturbatively
expandable about the $q=0$ theory with
the boundary condition \eqref{eq:periodic-q=0-bcs}~\cite{Simon:1990ic}. 

Second, if $q>0$ is fixed, $E_m^{\beta}$ is close to $E_m^{q=0, \beta}$ 
only for those $m$ for which $|2\pi m+ \beta| \ll 1/q$. 
In particular, for fixed $q>0$, the spectrum is unbounded both above and below, 
and the asymptotic behaviour of the large positive and negative 
eigenenergies is dominated by the $p^3$ term in the Hamiltonian. 
The perturbative expandability does hence not hold 
uniformly over the full set of the eigenvalues. 

Third, we may characterise the non-uniformity in the small $q$ 
behaviour in terms of the time 
evolution operator $V^{q,\beta}_t = \exp(-i\hbar^{-1} H^{q,\beta} t)$: 
it is straightforward to verify that for each $t$ and~$\beta$, 
$V^{q,\beta}_t$ converges to $V^{q=0,\beta}_t$ as $q\to0$ in the 
strong operator topology but not in the operator norm topology.

\subsection{Boundary conditions independent of $\phi''$\label{subsec:bc-nophipp}}

The boundary conditions for the unperturbed theory 
involve the values of the wave function and of its first derivative 
at the boundary, but not the values of the higher derivatives. 
We now ask: which boundary conditions for the perturbed theory 
involve only the values of the wave function 
and of its first derivative at the boundary? 

Requiring $\rho_3$ and $\sigma_3$ to drop out of~\eqref{eq:3by3condition}, and
using~\eqref{eq:ev-identity1}, we find that $U$ is given by {\small
\begin{align}
\frac{1}{\bigl(\sqrt{-\lambda_-}\, |a_3| \bigr)^2 }
\begin{pmatrix}
0 
&  \bigl(\sqrt{-\lambda_-}\, a_3 \bigr) \bigl(\sqrt{\lambda_+}\, b_3^*\bigr)
&  \bigl(\sqrt{-\lambda_-}\, a_3 \bigr) \bigl(\sqrt{\lambda_0}\, c_3^*\bigr)
\\[1ex]
\bigl(\sqrt{-\lambda_-}\, a_3^* \bigr) \bigl(\sqrt{\lambda_+}\, b_3\bigr)
& \bigl(\sqrt{\lambda_0}\, c_3\bigr)^2 \alpha 
& -  \bigl(\sqrt{\lambda_+}\, b_3\bigr) \bigl(\sqrt{\lambda_0}\, c_3\bigr) \alpha 
\\[1ex]
\bigl(\sqrt{-\lambda_-}\, a_3^* \bigr) \bigl(\sqrt{\lambda_0}\, c_3\bigr)
& -  \bigl(\sqrt{\lambda_+}\, b_3\bigr) \bigl(\sqrt{\lambda_0}\, c_3\bigr) \alpha 
&  \bigl(\sqrt{\lambda_+}\, b_3\bigr)^2 \alpha
\\
\end{pmatrix}
\ , 
\label{eq:U-noprimeprime}
\end{align}}%
where 
the only remaining
freedom is in the choice of the parameter $\alpha\in\BbbC$ with
$|\alpha|=1$. It can be verified, using \eqref{eq:ev-identity1} and
four other similar identities, that \eqref{eq:3by3condition} with $U$
given by \eqref{eq:U-noprimeprime} is equivalent to
\begin{subequations}
\label{eq:bc-elem-noprimeprime}
\begin{align}
&\phi(0) = \phi(L) = 0 
\ , 
\label{eq:bc-elem-noprimeprime-Dir}
\\
&\phi'(L) = e^{i\beta} \phi'(0)
\ , 
\label{eq:bc-elem-noprimeprime-quasiper}
\end{align}
\end{subequations}
where $0\le \beta < 2\pi$ and $e^{i\beta}$ is proportional to 
$\alpha$ in \eqref{eq:U-noprimeprime} by a phase that is determined 
by the phase choices of the
eigen-covectors~\eqref{eq:eigen-covectors}. 
(With the phase choices made in Appendix~\ref{app:expansion}, 
$e^{i\beta} = \alpha$.)

The main observation for us is that while the unperturbed Dirichlet wave 
functions satisfy \eqref{eq:bc-elem-noprimeprime-Dir}, 
they do not satisfy~\eqref{eq:bc-elem-noprimeprime-quasiper}, 
even though a subset of them satisfies 
\eqref{eq:bc-elem-noprimeprime-quasiper} 
for $\beta=0$ and the complementary subset for $\beta=\pi$. 
This means that 
the derivatives of the perturbed wave functions 
cannot converge to those in the unperturbed Dirichlet 
theory at least near the boundaries. Numerical
experiments suggest that as $q\to0$ with fixed~$\beta$, 
the low-lying positive eigenenergies do converge 
to the $q=0$ Dirichlet eigenenergies
$E_m^{q=0,\text{Dirichlet}}$~\eqref{eq:q=0-Dirichlet-eigen}; sample
numerical data is shown in Table~\ref{tab:Dirichlet-like}. 
This means that as $q\to0$, the perturbed eigenfunctions must contain 
a rapidly oscillating component, 
with the asymptotic
form $\exp\bigl(i x/(qL)\bigr)$, which plays an essential role in
satisfying~\eqref{eq:bc-elem-noprimeprime}. 
A~similar rapidly oscillating component can be 
verified to occur when the $p^3$ correction is replaced by a 
$p^4$ correction~\cite{louko-marples}. 

In summary, the boundary conditions independent 
of the second derivatives cannot 
yield a theory in which both the eigenenergies and the 
eigenfunctions are perturbatively expandable at small $q$  
in the sense that we are looking for. 

\begin{table}[t]
\centering
\begin{tabular}{l l | l l l l | l l l l}
& & \hbox to 0pt{$q = 10^{-2}$\hss} &&&& \hbox to 0pt{$q = 10^{-4}$\hss} \\[1ex]
\hline\hline
$m \vphantom{{A^A}^{A^A}_{{A_A}_{A_A}}}$ & $m^2$ 
& $\beta=0$ & $\beta=\frac12\pi$ & $\beta= \pi$  & $\beta=\frac32\pi$ 
& $\beta=0$ & $\beta=\frac12\pi$ & $\beta= \pi$  & $\beta=\frac32\pi$ \\
\hline
1 & 1 & 1.129& 1.030& 0.9997& 0.9440 & 1.00006& 1.00005& 1.00000& 1.00007 \\  
2 & 4 & 3.999& 3.924& 3.889& 3.792 &  4.00000& 3.99861& 3.98966& 4.00189 \\  
3 & 9 & 9.205& 9.079& 8.962& 8.023 & 9.00053& 9.00046& 9.00000& 9.00062 \\  
\vdots & \vdots & \vdots & \vdots & \vdots & \vdots & \vdots & \vdots
& \vdots & \vdots \\  
11 & 121 & 84.42& 80.03& 77.07& 75.37 & 121.001& 121.001& 120.999& 121.002 \\   
12 & 144 & 92.18& 90.36& 89.26& 87.76 & 143.999& 143.946& 143.320& 144.063 \\
\hline\hline
\end{tabular}
\caption{The table shows numerical results for ${(L/\pi)}^2$ times the
12 lowest positive eigenenergies for 
$q = 10^{-2}$ 
and 
$q = 10^{-4}$
under the boundary condition 
\eqref{eq:bc-elem-noprimeprime} with  
$\beta=0$, $\beta=\frac12\pi$, $\beta= \pi$, and $\beta=\frac32\pi$, 
enumerated by the index $m$ shown in the first column, and 
suppressing $m=4,5,\ldots,10$ where the pattern continues in a
straightforward way. 
The second column shows ${(L/\pi)}^2$ times the corresponding 
eigenenergies 
$E_m^{q=0, \text{Dirichlet}}$
\eqref{eq:q=0-Dirichlet-eigen}
in the $q=0$ theory with the Dirichlet boundary condition.  
The data suggests that the eigenenergies are converging to 
$E_m^{q=0, \text{Dirichlet}}$ as $q\to0$.}
\label{tab:Dirichlet-like}
\end{table}

For use in Section~\ref{sec:pert-spectrum}, we record here that when
the phases of the eigen-covectors \eqref{eq:eigen-covectors} are
chosen as in Appendix~\ref{app:expansion}, the matrix
\eqref{eq:U-noprimeprime} has the small $q$ expansion
\begin{align}
\begin{pmatrix}
0 
& -1+q-\tfrac12 q^2 + \cdots
&  \sqrt{2q}\,( 1 - \tfrac12 q + \cdots)
\\[1ex]
-1+q-\tfrac12 q^2 + \cdots
& 2q (1-q+ \cdots)\alpha
& \sqrt{2q} \, ( 1 - \tfrac32 q + \cdots )\alpha
\\[1ex]
\sqrt{2q} \, ( 1 - \tfrac12 q + \cdots )
&  \sqrt{2q}\, ( 1 - \tfrac32 q + \cdots )\alpha
& (1 - 2 q + 2q^2 + \cdots)\alpha
\\
\end{pmatrix}
\ . 
\label{eq:U-noprimeprime-smallq}
\end{align}

\section{Near-Dirichlet spectrum 
at small $q$\label{sec:pert-spectrum}}

We saw in subsection \ref{subsec:bc-nophipp} that the Dirichlet
condition of the $q=0$ theory 
does not generalise in a straightforward way to $q>0$, where 
any boundary condition 
that does not involve $\phi''$ must be in the 
$U(1)$ family~\eqref{eq:bc-elem-noprimeprime}. 
We now show that when $q$ is positive but small, there is a 
family of boundary conditions that are close to the 
$q=0$ Dirichlet theory 
in the sense of perturbative expandability of both the eigenenergies 
and the corresponding wave functions. 

We look for solutions to the eigenvalue equation $H\phi = E \phi$ in the form 
\begin{align}
\exp(i r_+ x/L) - B \exp(i r_- x/L)
\ , 
\label{eq:phi-pert-ansatz}
\end{align}
where $r_- = - m \pi \ + $ (corrections in~$q$) 
with $m = 1, 2, \ldots\,$,  
$r_+ = \bigl(1-q r_- - \sqrt{1+2 q r_- - 3 q^2 r_-^2} \, \bigr)/(2q) 
= m \pi  \ + $ (corrections in~$q$), 
and $B = 1 \ +$ (corrections in~$q$). 
The expression for $r_+$ in terms of $r_-$ comes from the eigenvalue equation, 
and $E = (r_-^2 - q r_-^3) L^{-2}$.  
When $q\to0$, \eqref{eq:phi-pert-ansatz} reduces to the 
$q=0$ Dirichlet eigenfunction \eqref{eq:q=0-Dir-efunctions}, 
and $E$ reduces to $E_m^{q=0, \text{Dirichlet}}$~\eqref{eq:q=0-Dirichlet-eigen}. 
Note that we have excluded from 
\eqref{eq:phi-pert-ansatz} a term proportional to 
the third linearly independent solution to $H\phi = E \phi$, 
given by $\exp(i r_0 x/L)$ where 
$r_0 = \bigl(1-q r_- + \sqrt{1+2 q r_- - 3 q^2 r_-^2} \, \bigr)/(2q)$, 
because $r_0$ diverges as $q\to0$ so that the wave function 
with this term present 
would not be perturbatively expandable in~$q$. 

We choose the phases of the eigen-covectors \eqref{eq:eigen-covectors} so that 
$a_1>0$, $b_1>0$ and $c_3>0$. From the small $q$ expansions given in Appendix \ref{app:expansion} 
it is seen that the $q=0$ Dirichlet condition is then obtained from \eqref{eq:3by3condition} 
by setting $q=0$ and 
\begin{align}
U = 
\begin{pmatrix}
0 & -1 & 0 \\
-1 & 0 & 0 \\
0 & 0 & - s \\
\end{pmatrix} 
\label{eq:U-without-U0}
\end{align}
where $s$ may be any complex number of unit modulus. 
We hence look for a $q>0$ boundary condition 
in which the matrix $U$ in \eqref{eq:3by3condition} 
has the form  
\begin{align}
U = 
\begin{pmatrix}
0 & -1 & 0 \\
-1 & 0 & 0 \\
0 & 0 & - s \\
\end{pmatrix} 
U_0
\ , 
\label{eq:U-ito-U0}
\end{align}
where $s$ is a $q$-independent complex number of unit modulus and the unitary matrix 
$U_0$ is the $3\times3$ identity matrix plus corrections in~$q$. 

The formulas in Appendix \ref{app:expansion} show that the $q$-dependent 
coefficients in \eqref{eq:3by3condition} have small $q$ expansions 
that proceed in positive integer powers of~$q^{1/2}$. 
We hence assume that $U_0$, $r_-$ and $B$ have expansions that proceed 
in positive integer powers~of~$q^{1/2}$. 
We find that there are exactly two ways to make the expansions consistent to order~$q^2$. These are as follows: 

{\bf Case I.} Set $U_0$ to the identity matrix and let $s$ remain arbitrary. To order~$q^2$, 
we then find 
\begin{subequations}
\label{eq:U-naive-expansions}
\begin{align}
r_- & = - m \pi + \tfrac12 {m}^{2}{\pi }^{2} q
\ , 
\\
r_+ & = m \pi + \tfrac12 {m}^{2}{\pi }^{2} q
\ , 
\\
B & = 1-m\pi q +\tfrac12 {m}^{2}{\pi }^{2} q^2 
\ , 
\\
E & = E_m^{\text{pert}} 
:= \left({m}^{2}{\pi }^{2} - \tfrac{5}{4}\,{m}^{4}{\pi }^{4} q^2 \right) L^{-2}
\ . 
\label{eq:E-naive-quadratic}
\end{align}
\end{subequations}
The correction in the eigenergies \eqref{eq:E-naive-quadratic} 
occurs in order~$q^2$, which is higher than one might 
have expected on grounds of the order $q$ term in~$H$. 
Note that none of the formulas in \eqref{eq:U-naive-expansions} 
depend on~$s$. 

{\bf Case II.}  Set $s=\pm1$ and 
\begin{align}
U_0 = \exp\left[
i \bigl (k_1 q^{1/2} + k_2 q + k_3 q^{3/2} + k_4 q^2 + O(q^{5/2}) \bigr)  
\begin{pmatrix}
-1 & s & 0 \\
s & -1 & 0 \\
0 & 0 & 2 \\
\end{pmatrix} 
\right]
\ , 
\label{eq:U0-pert-exp}
\end{align}
where $k_1$, $k_2$, $k_3$ and $k_4$
are real-valued constants, not all of them vanishing. 
The expressions \eqref{eq:U-naive-expansions} 
then acquire additional terms proportional to $q^{1/2}$, 
$q$, $q^{3/2}$ and~$q^2$, 
with coefficients that involve $s$ and positive powers of 
$k_1$, $k_2$, $k_3$ and~$k_4$. 
We record here only the expression for the eigenenergy: 
\begin{align}
E &= 
\biggl(
{m}^{2}{\pi }^{2} 
+ 2 m^{2}{\pi }^{2}k_1 \bigl(  (-1)^{m}s+1 \bigr) q^{1/2}
\notag
\\
& 
\hspace{5ex} 
+ 2{m}^{2}{\pi }^{2} 
\left( 3 k_1^{2} + k_2 \right)  \bigl(  (-1)^{m}s+1 \bigr) q 
\notag
\\
& 
\hspace{5ex} 
+ \tfrac23 {m}^{2} \pi^2
\left[
k_1^3 (25 - m^2 \pi^2)
+18 k_1 k_2 + 3 k_3 \right]  \bigl(  (-1)^{m}s+1 \bigr) q^{3/2}
\notag
\\
& 
\hspace{5ex} 
+ \Bigl\{ - \tfrac54
{m}^{4}{\pi }^{4}
+\tfrac23 m^2 \pi^2 
\bigl[ 2 k_1^4  ( 33 - 5 m^2 \pi^2 ) 
+ 3 k_1^2 k_2 ( 25 - m^2\pi^2 ) 
\notag
\\
& 
\hspace{10ex}
+9  k_2^{2}
+ 18 k_1  k_3 
+ 3 k_4 \bigr] 
\bigl(  (-1)^{m}s+1 \bigr)
\Bigr\} q^2 
\biggr) L^{-2}
\ . 
\label{eq:E-fourpar-quadratic}
\end{align}

Three comments are in order. First, note that Case I and Case II 
are distinct because in Case II we have assumed 
at least one of the constants $k_1$, $k_2$,
$k_3$ and $k_4$ to be nonvanishing. In the limit in which all four of
these constants are taken to zero, Case II reduces to Case I with $s=\pm1$.

Second, neither Case I nor Case II includes any of the $\phi''$-independent
boundary conditions~\eqref{eq:bc-elem-noprimeprime}. This can be seen
by comparing \eqref{eq:U-noprimeprime-smallq} to \eqref{eq:U-without-U0}
and to \eqref{eq:U-ito-U0} with~\eqref{eq:U0-pert-exp}.

Third, neither Case I nor Case II remains consistent when the perturbative
expansion is continued beyond order~$q^2$.

It should be emphasised that while we have shown that there exists a set of 
eigenenergies and eigenfunctions such that each of them 
is perturbatively expandable as $q\to0$, 
we have not examined whether there exists a sense
in which the expandability might hold uniformly over the full set of eigenenergies. 
The example of subsection \ref{subsec:q=0:per-upto-phase}
suggests that the spectrum for fixed $q>0$ is likely 
to be unbounded both from above and from below, 
the asymptotic behaviour of the large positive and 
negative eigenenergies to be dominated by the $p^3$ term in the Hamiltonian, 
and the expandability not to hold uniformly over the eigenenergy set. 
The closeness of the time evolution operator of the perturbed theory 
to that of the unperturbed theory might nevertheless again be 
characterisable in terms of convergence in an appropriate operator topology; 
however, verifying such convergence properties would require new techniques 
for analysing the full set of eigenenergies at fixed $q>0$. 

As a final comment, we note that numerical evidence, 
shown in Tables \ref{tab:q=0.01} and~\ref{tab:q=0.0001}, 
indicates that even within the energy range over which the
perturbative formula \eqref{eq:E-naive-quadratic} gives a good
approximation to the eigenenergies, there occur occasional
intercalating, nonperturbative eigenenergies that are not covered by
the perturbative formula. These nonperturbative eigenenergies
appear however to become rarer as $q$ decreases. 

\begin{table}[p]
\centering
\begin{tabular}{l|l | l l}
\multicolumn{4}{l}{$q = 10^{-2}$} \\[1ex]
\hline\hline
$m$ & $L^2 E_m^{\text{pert}}/\pi^2 \vphantom{{A^A}^{A^A}_{{A_A}_{A_A}}}$ & $s=1$  & $s=-1$  \\
\hline
1  & 0.9987663  & 0.9987654  & 0.9987658 \\
2  & 3.98026    & 3.98020    & 3.98021   \\
3  & 8.90007    & 8.89945    & 8.89950   \\
4  & 15.6842    & 15.6799    & 15.6808   \\
5  & 24.2289    & 24.2055    & 24.2145   \\
-- & --         & 25.0998    & --        \\
6  & 34.4011    & 34.3599    & 34.3597   \\
7  & 46.04      & 45.94      & 45.92     \\
-- & --         & --         & 51.74     \\
8  & 58.95      & 58.70      & 58.75     \\
9  & 72.91      & 72.16      & 72.41     \\
-- & --         & 74.98      & --        \\
10 & 87.66      & 86.77      & 86.59     \\
-- & --         & --         & 94.17     \\
11 & 102.9      & 101.1      & 101.3     \\
-- & --         & 110.4      & --        \\
12 & 118.4      & 115.2      & 115.2     \\
-- & --         & --         & 124.1     \\
13 & 133.8      & 127.8      & 128.3     \\
-- & --         & 134.2      & --        \\
14 & 148.7      & 140.0      & 138.7     \\
-- & --         & --         & 141.5     \\
15 & 162.5      & 146.3      & 146.7     \\
\hline\hline
\end{tabular}
\caption{$q = 10^{-2}$. 
The last two columns show numerical results for 
${(L/\pi)}^2$ times the 19 lowest positive eigenenergies 
under the boundary condition \eqref{eq:U-without-U0} 
with $s = \pm1$. 
The perturbative eigenenergies $E_m^{\text{pert}}$ 
\eqref{eq:E-naive-quadratic}, shown in the second column, 
provide a good approximation to 
15 of the eigenenergies, with $1\le m \le 15$,  
both for $s=1$ and for $s=-1$, to five decimal places near 
the lower end and to 10\%  near the upper end. 
For each of $s=1$ and $s=-1$, there are four  
eigenenergies are not close to $E_m^{\text{pert}}$, 
and these four nonperturbative eigenenergies 
intercalate between the perturbative 
ones differently for $s=1$ and $s=-1$.}
\label{tab:q=0.01}
\end{table}

\begin{table}[p]
\centering
\begin{tabular}{l|l | l l}
\multicolumn{4}{l}{$q = 10^{-4}$} \\[1ex]
\hline\hline
$m$ & $L^2 E_m^{\text{pert}}/\pi^2 \vphantom{{A^A}^{A^A}_{{A_A}_{A_A}}}$ & $s=1$  & $s=-1$  \\
\hline
1&
 0.999999876629944986 &
 0.999999876629936463 &
 0.999999876629936465 \\  
2&
 3.999998026079120 &
 3.999998026078571 &
 3.999998026078574 \\  
3&
 8.999990007025544 &
 8.999990007019330 &
 8.999990007019331 \\  
\vdots&\vdots&\vdots&\vdots\\  
17&
 288.9896960096 &
 288.9896958038 &
 288.9896958039 \\
 --&
 --& 
 314.6674566416 &
 --\\ 
18&
 323.9870491051 &
 323.9870488796 &
 323.9870488152 \\
\vdots&\vdots&\vdots&\vdots\\  
59&
 3479.505081 &
 3479.504721 &
 3479.504700 \\ 
 -- &
 -- & 
 -- &
 3495.371306 \\ 
60&
 3598.401124 &
 3598.400726 &
 3598.400731 \\
\vdots&\vdots&\vdots&\vdots\\  
81&
 6555.689324 &
 6555.686871 &
 6555.686914 \\  
-- &
-- & 
 6672.077214 &
 --\\  
82&
 6718.422171 &
 6718.419579 &
 6718.419577 \\
\vdots&\vdots&\vdots&\vdots\\  
99&
 9789.149121 &
 9789.141082 &
 9789.140921 \\  
 -- &
 -- &
 -- &
 9844.787186 \\  
100&
 9987.662994 &
 9987.654463 &
 9987.654503 \\
\hline\hline
\end{tabular}
\caption{$q = 10^{-4}$. 
As in Table \ref{tab:q=0.01}, for the 102 
lowest positive eigenenergies, suppressing the ranges of $m$ where the pattern
continues in a straightforward way. Apart from the two nonperturbative 
eigenenergies for each~$s$, 
$E_m^{\text{pert}}$ \eqref{eq:E-naive-quadratic} 
is accurate to 11 decimal places near the lower end and to five decimal 
places near the upper end.} 
\label{tab:q=0.0001}
\end{table}


\section{Conclusions\label{sec:conclusions}}

We have discussed the quantum mechanics of a nonrelativistic 
particle on an interval of finite length when the Hamiltonian contains 
a correction term proportional to~$p^3$. 
We gave an explicit description of the $U(3)$ 
family of self-adjoint extensions of the Hamiltonian, and we showed that 
the only boundary conditions that do not involve the second 
derivative of the wave function
require the wave function to vanish at the two ends and its derivative 
to be equal at the two ends up to a prescribed phase. 
This implies in particular that 
the Dirichlet condition of setting the 
wave function to zero at the two ends 
does not qualify on its own as a self-adjointness condition. 

We saw that periodicity up to a prescribed phase does belong to the
$U(3)$ family of self-adjointness conditions. The eigenenergies and
eigenfunctions were written down in terms of elementary expressions,
and we noted that both the eigenenergies and the eigenfunctions are
perturbatively expandable about the limit in which the coefficient of
the $p^3$ correction term vanishes. 
While the expandability is not uniform 
over the eigenvalue set, it is sufficiently strong to 
make the time evolution operator of the perturbed theory 
converge to that of the unperturbed theory 
in the strong operator topology, 
although not in the operator norm topology. 

Our main result was to find a subfamily of self-adjointness
conditions, indexed by four continuous parameters and one binary
parameter, under which there exists a countable set of eigenenergies
and corresponding eigenfunctions that are perturbatively close to
those of the uncorrected nonrelativistic particle under Dirichlet
boundary conditions.  We further showed that this subfamily is unique,
subject to certain technical assumptions. The closeness holds
individually for each of the perturbative eigenvalues, but we did not
attempt to give the closeness a sense that would be valid uniformly
over the full set of eigenvalues. 
It might be possible to characterise this closeness in terms of the topology 
in which the perturbed time evolution operator converges 
to the unperturbed one, but such an analysis would 
require a better control over the global properties of the perturbed spectrum. 

The physical motivation to consider a Hamiltonian with the $p^3$
correction term was that this term may model low energy effects due to
quantum gravity~\cite{AmelinoCamelia:2008qg}. Our main result
shows that the quantum theory in the presence of this term can be
formulated on the interval so as to be unitary and perturbatively
close to the uncorrected particle with the Dirichlet boundary
conditions. The special interest of the Dirichlet conditions here is
that they can be regarded as generic in the uncorrected theory when
the two ends of the interval are considered to be independent of each
other~\cite{walton}.

Finally, we saw that the eigenenergies in our near-Dirichlet theories
depend on the coefficient of the $p^3$ term through positive integer
and half-integer powers, without rapid oscillations or other signs of
irregularity. Our near-Dirichlet boundary conditions hence do not
single out for this coefficient discrete values that could be regarded
as a quantisation condition on the length of the interval in terms of
the underlying quantum gravity scale~\cite{Ali:2009zq}.

\section*{Acknowledgments}

I thank Saurya Das and Elias Vagenas for helpful correspondence
and an anonymous referee for helpful comments. 
This work was supported in part by STFC (Theory
Consolidated Grant ST/J000388/1).


\begin{appendix}

\section{Appendix: Subspaces of self-adjointness\label{app:subspaces}}

In this appendix we perform the maximal linear subspace analysis that 
leads to the self-adjointness boundary conditions \eqref{eq:3by3condition} 
in the main text. 

\subsection{Preliminaries}

Let $n$ be a positive integer and $\mathcal{H} = \BbbC^{2n}$. Define
on $\mathcal{H}$ the Hermitian form 
\begin{align}
B(u,v) = u^\dagger 
\begin{pmatrix}
I & 0 \\
0 & -I
\end{pmatrix}
v 
\ , 
\end{align}
where $I$ is the $n\times n$ identity matrix. 

{\bf Lemma.}\ 
The maximal linear subspaces $V \subset \mathcal{H}$ 
on which $B(u,v)=0$ for all $u,v \in V$ are 
\begin{align}
V_U = \bigl\{ v \in
\mathcal{H} \mid 
\bigl(
\begin{smallmatrix}
U & -I \\
0 & 0
\end{smallmatrix}
\bigr) 
v =0
\bigr\}
\ , 
\label{eq:UV-in-lemma}
\end{align}
where $U \in U(n)$. 

\emph{Proof.}\ 
Let $V \subset \mathcal{H}$ be a linear subspace 
on which $B(u,v)=0$ for all $u,v \in V$. Suppose  
$w = \bigl(
\begin{smallmatrix}
w_1 \\ w_2 
\end{smallmatrix}
\bigr) \in V$ where $w_1, w_2 \in \BbbC^n$.  
Then $B(w,w)=0$ implies $\Vert w_1\Vert = \Vert w_2\Vert$. 
As $V$ is a linear subspace, 
each $v \in V$ must hence have the form 
$\bigl(
\begin{smallmatrix}
v_1 \\ U v_1 
\end{smallmatrix}
\bigr)$, where $U$ is a constant $n\times n$ matrix, such that 
if $V_1 \subset
\BbbC^n$ denotes the projection of $V$ to its first $n$ components,
$U$ maps $V_1$ isometrically to~$\BbbC^n$.  
For $u = \bigl(
\begin{smallmatrix}
u_1 \\ U u_1
\end{smallmatrix}
\bigr)$ 
and 
$v = \bigl(
\begin{smallmatrix}
v_1 \\ U v_1 
\end{smallmatrix}
\bigr)$ in $V$, $B(u,v)=0$ is equivalent to 
$u_1^\dagger \bigl(U^\dagger U - I  \bigr) v_1 = 0$. 
This holds for all $u_1 , v_1 \in  \BbbC^n$ iff 
$U^\dagger U = I$. $\blacksquare$ 

{\bf Remark.}\ 
The maximal linear subspaces on which 
$B(v,v)=0$ coincide with~\eqref{eq:UV-in-lemma}. 
The proof is as above but setting at every step $u=v$.

For generalisations, see \cite{kochubei,bruening}.

\subsection{Main proposition}

Let $n$ be a positive integer and $\mathcal{H} = \BbbC^{2n}$. 
Define on $\mathcal{H}$ the Hermitian form 
\begin{align}
C(u,v) = u^\dagger A v 
\ , 
\label{eq:Cform-def}
\end{align}
where $A$ is a Hermitian $2n\times 2n$ matrix with $n$ strictly
positive eigenvalues and $n$ strictly negative eigenvalues (each
eigenvalue counted with its multiplicity). 
By matrix diagonalisation, 
there exists a unitary
$2n\times 2n$ matrix $P$ and a real diagonal positive definite 
$2n\times 2n$
matrix $D$ such that 
\begin{align}
A = {(D P)}^\dagger 
\begin{pmatrix}
I& 0 \\
0 & -I  
\end{pmatrix} 
(D P)
\ . 
\label{eq:A-DPconjugate}
\end{align}

{\bf Proposition.}\
The maximal linear subspaces $V \subset \mathcal{H}$ 
on which $C(u,v)=0$ for all $u,v \in V$ 
are 
\begin{align}
V_U = \bigl\{ v \in
\mathcal{H} \mid 
\bigl(
\begin{smallmatrix}
U & -I \\
0 & 0
\end{smallmatrix}
\bigr) (D P) v =0
\bigr\}
\ , 
\label{eq:sa-conditions}
\end{align}
where $U \in U(n)$. 

\emph{Proof.}\ 
Follows from the Lemma by observing that 
$C(u,v) = B( DP u, D P v)$. 
$\blacksquare$ 

\subsection{Application\label{subsec:specialcase}}

We specialise \eqref{eq:Cform-def} to 
\begin{align}
A = 
\begin{pmatrix}
G& 0 \\
0 & -G  
\end{pmatrix}
\label{eq:A-intermsof-G} 
\end{align}
where $G$ is a Hermitian $3\times3$ matrix with the eigenvalues 
$\lambda_-<0$, $\lambda_+>0$ and $\lambda_0>0$ and the corresponding
orthogonal normalised eigen-covectors 
\begin{align}
\begin{pmatrix}
a_1& a_2 & a_3
\end{pmatrix} 
\ , 
\ \ 
\begin{pmatrix}
b_1& b_2 & b_3
\end{pmatrix} 
\ , 
\ \ 
\begin{pmatrix}
c_1& c_2 & c_3
\end{pmatrix} 
\ . 
\label{eq:eigen-covectors-app} 
\end{align}
The matrix
\begin{align}
\tilde P 
= 
\begin{pmatrix}
a_1& a_2 & a_3 \\
b_1& b_2 & b_3 \\ 
c_1& c_2 & c_3 \\
\end{pmatrix} 
\end{align}
is then unitary, and 
\begin{align}
\begin{pmatrix}
\tilde P& 0 \\
0 & \tilde P  
\end{pmatrix} 
A 
\begin{pmatrix}
\tilde P^\dagger& 0 \\
0 & \tilde P^\dagger 
\end{pmatrix} 
= 
\diag (\lambda_-, \lambda_+, \lambda_0,- \lambda_-, - \lambda_+, - \lambda_0)
\ . 
\end{align}
Let 
\begin{align}
Q &= 
\begin{pmatrix}
0& 0 & 0 & 1 & 0 & 0 \\
0& 1 & 0 & 0 & 0 & 0 \\
0& 0 & 1 & 0 & 0 & 0 \\
1& 0 & 0 & 0 & 0 & 0 \\
0& 0 & 0 & 0 & 1 & 0 \\
0& 0 & 0 & 0 & 0 & 1 \\
\end{pmatrix} 
\ ,
\\[2ex]
\tilde D & = 
\begin{pmatrix}
\sqrt{-\lambda_-}& 0 & 0  \\
0& \sqrt{\lambda_+} & 0 \\
0& 0 & \sqrt{\lambda_0} \\
\end{pmatrix} 
\ . 
\end{align}
Then 
\eqref{eq:A-DPconjugate}
holds with 
\begin{align}
DP = 
\begin{pmatrix}
\tilde D& 0 \\
0 & \tilde D  
\end{pmatrix} 
Q 
\begin{pmatrix}
\tilde P& 0 \\
0 & \tilde P  
\end{pmatrix} 
\ . 
\end{align}
Writing in \eqref{eq:sa-conditions}
\begin{align}
v = 
\begin{pmatrix}
\rho_1 \\
\rho_2 \\
\rho_3 \\
\sigma_1 \\
\sigma_2 \\
\sigma_3 \\
\end{pmatrix} 
\ , 
\end{align}
we have 
\begin{align}
DP 
\begin{pmatrix}
\rho_1 \\
\rho_2 \\
\rho_3 \\
\sigma_1 \\
\sigma_2 \\
\sigma_3 \\
\end{pmatrix} 
= 
\begin{pmatrix}
\sqrt{-\lambda_-} \left( a_1 \sigma_1 + a_2 \sigma_2 + a_3 \sigma_3\right) \\
\sqrt{\lambda_+} \left( b_1 \rho_1 + b_2 \rho_2 + b_3 \rho_3\right) \\
\sqrt{\lambda_0} \left( c_1 \rho_1 + c_2 \rho_2 + c_3 \rho_3\right) \\
\sqrt{-\lambda_-} \left( a_1 \rho_1 + a_2 \rho_2 + a_3 \rho_3\right) \\
\sqrt{\lambda_+} \left( b_1 \sigma_1 + b_2 \sigma_2 + b_3 \sigma_3\right) \\
\sqrt{\lambda_0} \left( c_1 \sigma_1 + c_2 \sigma_2 + c_3 \sigma_3\right) \\
\end{pmatrix} 
\ , 
\end{align}
and the subspace condition \eqref{eq:sa-conditions} reads 
\begin{align}
U 
\begin{pmatrix}
\sqrt{-\lambda_-} \left( a_1 \sigma_1 + a_2 \sigma_2 + a_3 \sigma_3\right) \\
\sqrt{\lambda_+} \left( b_1 \rho_1 + b_2 \rho_2 + b_3 \rho_3\right) \\
\sqrt{\lambda_0} \left( c_1 \rho_1 + c_2 \rho_2 + c_3 \rho_3\right) \\
\end{pmatrix} 
= 
\begin{pmatrix}
\sqrt{-\lambda_-} \left( a_1 \rho_1 + a_2 \rho_2 + a_3 \rho_3\right) \\
\sqrt{\lambda_+} \left( b_1 \sigma_1 + b_2 \sigma_2 + b_3 \sigma_3\right) \\
\sqrt{\lambda_0} \left( c_1 \sigma_1 + c_2 \sigma_2 + c_3 \sigma_3\right) \\
\end{pmatrix}
\ . 
\label{eq:3by3condition-app}
\end{align}
This is the condition \eqref{eq:3by3condition} in the main text.

\section{Appendix: Small $q$ expansions of the eigenvalues 
and eigen-covectors\label{app:expansion}} 

In this appendix we give the small $q$ expansions of the eigenvalues 
and $\sqrt{|\lambda|}$ 
times the normalised eigen-covectors \eqref{eq:eigen-covectors} 
of the matrix~\eqref{eq:Gmatrix-cubic}. 
The phases of the eigen-covectors are chosen so that 
$a_1>0$, $b_1>0$ and $c_3>0$. 

\begin{subequations}
\begin{align}
\lambda_- & = 
-1+\frac12\,q-\frac58\,{q}^{2}-\frac12\,{q}^{3}-{\frac {7}{128}}\,{q}^{4}+\frac12\,{q}
^{5}+{\frac {675}{1024}}\,{q}^{6}
+ O({q}^8)
\\[1ex]
\lambda_+ & = 1+\frac12\,q+\frac58\,{q}^{2}-\frac12\,{q}^{3}+{\frac {7}{128}}\,{q}^{4}+\frac12\,{q}^
{5}-{\frac {675}{1024}}\,{q}^{6}
+ O({q}^8)
\\[1ex]
\lambda_0 & = 
{q}^{3} \left(1-{q}^{2}+3\,{q}^{6}
+ O({q}^8)
\right)
\end{align}
\end{subequations}
\begin{subequations}
\label{eq:denscovec-pert-minus}
\begin{align}
\sqrt{-\lambda_-} \,a_1 & = 
\frac{1}{\sqrt {2}} \Bigl( 1+\frac{3}{16}\,{q}^{2}-{\frac {83}{512}}\,{q}^{4}+{
\frac {3605}{8192}}\,{q}^{6}+\frac12\,{q}^{7}
+ O({q}^8)
\Bigr)
\\[1ex]
\sqrt{-\lambda_-} \,a_2 & = 
-\frac{i}{\sqrt{2}} \Bigl(
1 
-\frac12\,q
-\frac{3}{16}\,{q}^{2}
-{\frac {3}{32}}\,{q}^{3}
+{\frac {101}{512}}\,{q}^{4}
+{\frac {595}{1024}}\,{q}^{5}
+{\frac {4035}{8192}}\,{q}^{6}
\notag
\\[1ex]
& \hspace{11ex}
-{\frac {10261}{16384}}\,{q}^{7}
+ O({q}^8)
\Bigr)
\\[1ex]
\sqrt{-\lambda_-} \,a_3 & = 
\frac{q}{\sqrt{2}} \Bigl( 1+\frac12\,q-\frac{3}{16}\,{q}^{2}-{\frac {29}{32}}\,{q}^{3}-{\frac 
{411}{512}}\,{q}^{4}+{\frac {749}{1024}}\,{q}^{5}+{\frac {21955}{8192}
}\,{q}^{6}
\notag
\\[1ex]
& \hspace{11ex}
+{\frac {33909}{16384}}\,{q}^{7}
+ O({q}^8)
\Bigr)
\end{align}
\end{subequations}
\begin{subequations}
\label{eq:denscovec-pert-plus}
\begin{align}
\sqrt{\lambda_+} \,b_1 & = 
\frac{1}{\sqrt{2}}
\Bigl( 1+\frac{3}{16}\,{q}^{2}-{\frac {83}{512}}\,{q}^{4}+{\frac {3605}{
8192}}\,{q}^{6}-\frac12\,{q}^{7}
+ O({q}^8)
\Bigr) 
\\[1ex]
\sqrt{\lambda_+} \,b_2 & = 
\frac{i}{\sqrt{2}} 
\Bigl(
1
+\frac12\,q
-\frac{3}{16}\,{q}^{2}
+{\frac {3}{32}}\,{q}^{3}
+{\frac {101}{512}}\,{q}^{4}
-{\frac {595}{1024}}\,{q}^{5}
+{\frac {4035}{8192}}\,{q}^{6}
\notag
\\[1ex]
& \hspace{11ex}
+{\frac {10261}{16384}}\,{q}^{7}
+ O({q}^8)
\Bigr)
\\[1ex]
\sqrt{\lambda_+} \,b_3 & = 
-\frac{q}{\sqrt{2}} 
\Bigl( 
1-\frac12\,q-\frac{3}{16}\,{q}^{2}+{\frac {29}{32}}\,{q}^{3}-{
\frac {411}{512}}\,{q}^{4}-{\frac {749}{1024}}\,{q}^{5}+{\frac {21955}
{8192}}\,{q}^{6}
\notag
\\[1ex]
& \hspace{11ex}
-{\frac {33909}{16384}}\,{q}^{7}
+ O({q}^8)
\Bigr)
\end{align}
\end{subequations}
\begin{subequations}
\label{eq:denscovec-pert-nought}
\begin{align}
\sqrt{\lambda_0} \,c_1 & = 
-{q}^{7/2} \left( 1-2\,{q}^{2}+{q}^{4}+7\,{q}^{6}
+ O({q}^8)
\right) 
\\[1ex]
\sqrt{\lambda_0} \,c_2 & = 
i{q}^{5/2} \left( 1-{q}^{2}-{q}^{4}+7\,{q}^{6}
+ O({q}^8)
\right) 
\\[1ex]
\sqrt{\lambda_0} \,c_3 & = 
{q}^{3/2} \left( 
1-{q}^{2}+4\,{q}^{6}
+ O({q}^8)
\right) 
\end{align}
\end{subequations}

\end{appendix}


\begin{thebibliography}{99}

\bibitem{AmelinoCamelia:2008qg} 
  G.~Amelino-Camelia,
  ``Quantum-spacetime phenomenology,''
  Living Rev.\ Rel.\  {\bf 16}, 5 (2013)
  [arXiv:0806.0339 [gr-qc]].

\bibitem{Kempf:1994su} 
  A.~Kempf, G.~Mangano and R.~B.~Mann,
  ``Hilbert space representation of the minimal length uncertainty relation,''
  Phys.\ Rev.\ D {\bf 52}, 1108 (1995)
  [arXiv:hep-th/9412167].

\bibitem{Husain:2012im} 
  V.~Husain, D.~Kothawala and S.~S.~Seahra,
  ``Generalized uncertainty principles and quantum field theory,''
  Phys.\ Rev.\ D {\bf 87}, 025014 (2013)
  [arXiv:1208.5761 [hep-th]].

\bibitem{Ali:2009zq} 
A.~F.~Ali, S.~Das and E.~C.~Vagenas,
``Discreteness of space from the generalized uncertainty principle,''
Phys.\ Lett.\ B {\bf 678}, 497 (2009)
[arXiv:0906.5396 [hep-th]].

\bibitem{reed-simonII}
M.~Reed and B.~Simon, 
{\it Methods of Modern Mathematical Physics II: 
Fourier Analysis, Self-adjointness\/}
(Academic, New York, 1975).

\bibitem{blabk}
J.~Blank, 
P.~Exner
and 
M.~Havl\'{i}\v{c}ek, 
{\it Hilbert Space Operators in Quantum Physics\/}, 2nd edition
(Springer, New York, 2008). 

\bibitem{Bonneau:1999zq} 
G.~Bonneau, J.~Faraut and G.~Valent,
``Selfadjoint extensions of operators and the teaching of quantum mechanics,''
Am.\ J.\ Phys.\  {\bf 69}, 322 (2001)
[arXiv:quant-ph/0103153].

\bibitem{Asorey:2004kk} 
M.~Asorey, A.~Ibort and G.~Marmo,
``Global theory of quantum boundary conditions and topology change,''
Int.\ J.\ Mod.\ Phys.\ A {\bf 20}, 1001 (2005)
[arXiv:hep-th/0403048].

\bibitem{Ibort:2013ab} 
A.~Ibort and J.~M.~P\'erez-Pardo,
``Numerical solutions of the spectral problem for 
arbitrary self-adjoint extensions of the one-dimensional Schr\"odinger equation,'' 
SIAM J.\ Numer.\ Anal.\ {\bf 51}, 1254 (2013). 

\bibitem{Asorey:2013wca} 
M.~Asorey and J.~M.~Mu\~noz-Casta\~neda,
``Attractive and repulsive Casimir vacuum energy with general 
boundary conditions,''
Nucl.\ Phys.\ B {\bf 874}, 852 (2013)
[arXiv:1306.4370 [hep-th]].

\bibitem{Asorey:2013wvh} 
M.~Asorey, A.~P.~Balachandran and J.~M.~P\'erez-Pardo,
``Edge States: Topological Insulators, Superconductors and QCD Chiral Bags,''
JHEP {\bf 1312}, 073 (2013)
[arXiv:1308.5635 [cond-mat.mtrl-sci]].

\bibitem{Munoz-Castaneda:2014yea} 
J.~M.~Mu\~noz-Casta\~neda, K.~Kirsten and M.~Bordag,
``QFT over the finite line. Heat kernel coefficients, 
spectral zeta functions and selfadjoint extensions,''
Lett.\ Math.\ Phys.\ {\bf 105}, 523 (2015)
[arXiv:1402.7176 [math-ph]].

\bibitem{Balasubramanian:2014pba} 
V.~Balasubramanian, S.~Das and E.~C.~Vagenas,
``Generalized uncertainty principle and self-adjoint operators,''
arXiv:1404.3962 [hep-th].

\bibitem{walton} 
B.~Belchev and M.~A. Walton, 
``Robin boundary conditions and the Morse potential in quantum mechanics'', 
J.\ Phys.\ A {\bf43}, 085301 (2010)
[arXiv:1002.2139 [quant-ph]]. 

\bibitem{Simon:1990ic} 
J.~Z.~Simon,
``Higher-derivative Lagrangians, nonlocality, problems, and solutions,''
Phys.\ Rev.\ D {\bf 41}, 3720 (1990).

\bibitem{louko-marples}
J.~Louko and C.~R. Marples, unpublished (2014). 

\bibitem{kochubei}
A.~N. Kochubei, 
``Extensions of symmetric operators and symmetric binary
relations,''
Math.\ Notes {\bf 17}, 25 (1975). 

\bibitem{bruening}
J.~Br\"uning, 
V.~Geyler 
and 
K.~Pankrashkin, 
``Spectra of self-adjoint extensions 
and applications to solvable Schr\"odinger operators,'' 
Rev.\ Math.\ Phys.\ {\bf 20}, 1 (2008). 


\end{thebibliography}
\end{document}